\documentclass[conference]{IEEEtran}

\usepackage{mathptmx}       
\usepackage{helvet}         
\usepackage{courier}        
\usepackage{type1cm}        
\usepackage{makeidx}         
\usepackage{graphicx}        
\usepackage{multicol}        
\usepackage[bottom]{footmisc}

\usepackage{cite}
\usepackage{amsmath,amssymb,amsfonts}
\usepackage{algorithmic}
\usepackage{graphicx}
\usepackage{textcomp}

\begin{document}
\title{Collaborative SQL-Injections Detection System with Machine Learning}

\author{\IEEEauthorblockN{Mois\'{e}s Lodeiro-Santiago, C\'{a}ndido Caballero-Gil, Pino Caballero-Gil} \IEEEauthorblockA{Univesity of La Laguna, Avda. Astrofísico Francisco S\'{a}nchez s/n., San Crist\'{o}bal de La Laguna, Santa Cruz de Tenerife, \\ mlodeirs@ull.edu.es, ccabgil@ull.edu.es, pcaballe@ull.edu.es}}

\maketitle


\begin{abstract}

Data mining and information extraction from data is a field that has gained relevance in recent years thanks to techniques based on artificial intelligence and use of machine and deep learning. The main aim of the present work is the development of a tool based on a previous behaviour study of security audit tools (oriented to SQL pentesting) with the purpose of creating testing sets capable of performing an accurate detection of a SQL attack. The study is based on the information collected through the generated web server logs in a pentesting laboratory environment. Then, making use of the common extracted patterns from the logs, each attack vector has been classified in risk levels (dangerous attack, normal attack, non-attack, etc.). Finally, a training with the generated data was performed in order to obtain a classifier system that has a variable performance between 97 and 99 percent in positive attack detection. The training data is shared to other servers in order to create a distributed network capable of deciding if a query is an attack or is a real petition and inform to connected clients in order to block the petitions from the attacker's IP.

\end{abstract}

%

\section{Introduction}
\label{sec:introduccion}
Computer security is a topic of growing interest in the society due to the need to have secure systems that take care of information and protect it from "hackers". It is usual to find news in different media (TV, newspaper, Internet..) related to attacks on large companies \cite{wannacry} that had vulnerable systems. Nowadays, company projects or plans usually require a system where the information must be stored, which is why they make use of database engines that are often unprotected or using the default and unsafe configuration. For years, \cite{owasp} SQL attacks have allowed information stealing due to logical failures and a bad filter in the parameters and a non-validation of whether a query can be executed or not. This type of techniques is not only to the knowledge of the experts. In spite of being something technical, it is common to find new vulnerabilities day in day out in places like Packet Storm \cite{packetstorm} and Exploit DB \cite{exploitdb} to reach and execute any exploit (vulnerability) with a few clicks.

From a time there are tools that try to solve these problems. One of them is AMNESIA \cite{halfond2005amnesia} combining routine analysis and monitoring for a model construction of different types of real queries from an application. Queries that do not match the generated patterns are prevented from access. In \cite{buehrer2005using} it is used a classifier tree based on tokens extracted from some SQL queries and try to give a solution to incoming queries to check whether it is an attack or not. However, possibly due to the year of publishing, no cases are contemplated as a possible bypass in the queries (WAF evasion techniques) like including comments, hex transformation, etc.

In addition to this, several tools have been presented over the last years based on the use of Machine Learning techniques as a method to classify and evade intrusion in systems. Among the most cited cases it is \cite{7064617}, proposing a detection SQL-based attacks system based on the experience of the developers. This information is helpful for a few cases but is not specific or determinant today to avoid certain attacks. Furthermore, tokens-based totality analysis is usually not very reliable because a non-offensive query can contain statements like $where, or, and, etc.$. As an improvement, our system has been based on the heuristic and behaviour of SQL injection patterns starting from the scratch to ending with a test laboratory where the isolated information server was attacked by one of the most prestigious and well-known tools for computer audit in SQL engines, the SQLMap \cite{sqlmap}. In \cite{tajpour2011sql} a classification and analysis of the consequences of an attack by SQLIA as thefts in systems of authorization, alteration of the confidentiality and integrity are carried out. It also offers a rating based on attack attempts to determine whether it is a modification query, extraction query, schema definition, etc. Additionally, there are other projects of similar characteristic like \cite{6163109} and \cite{6041963} where an analysis of many types of web-oriented attacks (including SQL engine attacks) based on the use of attack vectors to identify SQL injections. The size of the used dataset for the classification it is quite small even for query pattern detection ($ <1000 $). On the contrary, in our project, the number of tests in the training dataset and the concept tests used in this paper (reaching a reliability of approximately 99 \%) reaches 80, 208 requests (where more than 75 \% are SQL queries). In \cite{sonoda2016approximate}, Michio proposes a statistical system close to how the statistical systems behind machine learning could work, using vector classifier systems with stochastic methods with approximation methods in parameters.

All solutions discussed above are based on the learning machine to distinguish actual requests for malicious requests that make use (usually) of attack vector analysis. However, none take into account factors such as the evolution of attack patterns, the obsolescence of detection systems or training time and the improvement of classifiers. This paper is based on the hypothesis that there is a network of machines (computers) distributed on the Internet where each computer can be labelled as a client, server or both at the same time. This network shares the information of the logs generated by the clients in order to be analysed by the servers to classify possible malign requests. Also, if a client is affected, the other clients can be protected against that menace using the information offered by the analysis servers. The main aim is to create a prototype based on emerging technologies such as FireBase \cite{firebase} (created by Google Inc.) as an instant connection socket between clients and servers allowing rapid reaction and propagation by the classifier servers.

The paper is structured as follows: In the \ref{sec:lab} section, an analysis of how the data have been collected in an empirical way is done. In section \ref{sec:col}, it is shown how a distributed system works and discusses an application of these distributed systems for SQLIA detection. Finally, the work ends with the conclusions \ref{sec:conc} where the future ideas of the current work are also shown.

\section{Laboratory and Data Extraction}
\label{sec:lab}

\begin{figure}[ht!]
	\includegraphics[width=\linewidth]{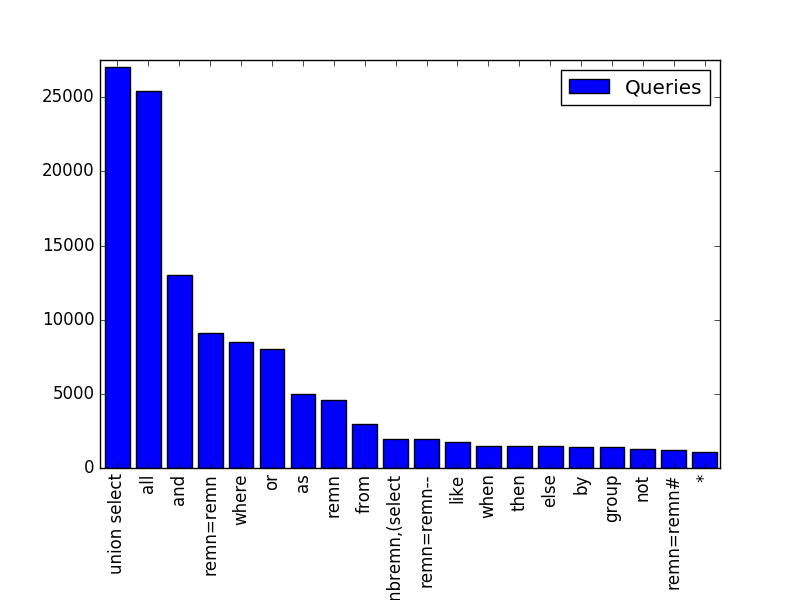}
    \caption{SQLIA Frequency attacks}
    \label{fig:figura_frec1}
\end{figure}

\begin{figure}[ht!]
	\includegraphics[width=\linewidth]{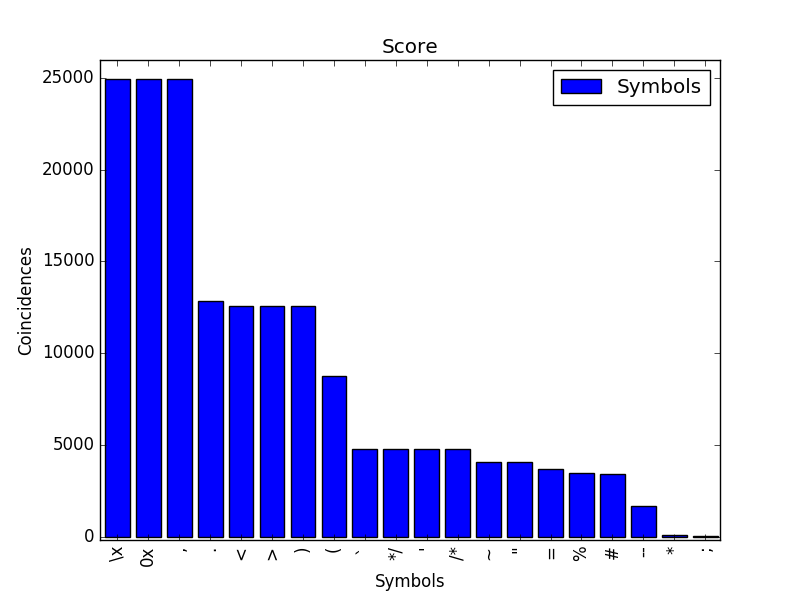}
    \caption{Frequency non-alphabetic analysis in SQLIA}
    \label{fig:figura_frec2}
\end{figure}

In the cited papers in the introduction, a reference is made on several occasions to the fact that the source of information on the extraction of knowledge is based on the experience of researchers. There is a lack of a scientific basis so, for the present study, an ad-hoc laboratory has been done with the single purpose of seeing the behaviour of some of the most usual database audit tools. In particular, a private laboratory with two machines has been carried out; The first machine is one with an Apache server with the logs enabled, the another machine (made up with docker) is a simple virtualised environment with the SQL Map tool installed. SQLMap tool is one of the most used tools for the detection of SQL vulnerabilities  since it allows the detection and even exploitation of vulnerabilities. The tests were performed with the following query: $sqlmap -u "http://IP/?id=1" --level=5 -risk=3 -forms -v 5 -threads=2 -random-agent -parse-errors$ in order to obtain the real records used by the tool. The tool is setted with level=5 and risk=3 (maximum values) and tests are performed from simple inspections like $'or 1 = 1--'$ to more complex injection attempts such as $uNiOn/**/all \\x0 \cdots$ and so on. This generates in the Apache server a record log file with a sum of 42,664 queries (malicious injections). For the creation of "legal" queries have been used records of several servers with different types of application (WordPress among them) with a total of approximately 8000 queries.

\subsection{Preparing queries and creating a dictionary}
In order to perform an automatic classification, a detection of patterns and tokens based on an analysis of raw file frequencies was done to divide the alphabetic characters $Payload1$ \ref{fig:figura_frec1} and separate the non-alphabetic characters $Payload2$ \ref{fig:figura_frec2}. After the data frequencies study, it has proceeded to the classification into 4 different types (1 without risk, 2 small risk, 3 risk and 4 high risk). Each sort of classification is based on the weighted use of the extracted elements in the analysis where, for instance, a level 4 (maximum) is equivalent to the use of a $union all$ statement. Non-alphabetic complements modify the previous level by adding $+1$ to the level in case of detection of common elements used in injections $\\x, 0x, <, >, etc$. The algorithm followed to create the dictionaries is the following one:

\begin{enumerate}
\item To start, using the list generated by SQLMap. All queries are marked as dangerous
\item $ilevel = 0, l = log\_file, s = symbol\_file$
\item $\forall l_x \subset l$ and $\forall p \subset Payload1$ $/ l_x = minus(l_x)$
  \begin{enumerate}
  \item  $l1 \rightarrow remove\_spaces(minus(l1))$ 
  \item $l2 \rightarrow htmlDecode(remove\_spaces(only\_chars(l1)))$
  \end{enumerate}
\item $if (Payload1 in l1) or (Payload2 in l2):$
  \begin{enumerate}
    \item $if union select \subset Payload1: level \rightarrow 4$
    \item $if level < 4 and Payload1 \subset \{all, chr, and, =, where, or\}: level \rightarrow 3$
    \item $if level < 3 and Payload1 \subset \{select, as, from, like, ...\}: level \rightarrow 2$
  \end{enumerate}
\item $if Payload in s:$
\begin{itemize}
	\item $if level < 4 and Payload1 in \{\\x, 0x, ', "\}: level \rightarrow +1$
    \item $if level < 3 and Payload1 in \{., <, >, (, ), number\}: level \rightarrow +1$
    \item $if level < 2 and Payload1 in \{\/*, *\/, \%, \#, --, ;\}: level \rightarrow +1$
\end{itemize}
\end{enumerate}

Once the classification process is finished, we have a vector $\vec{x}$ with 51 variables. The classification is represented by the last variable. The 30 first variables in $\vec{x}[0:30]$ represents the 30 most common patterns (see Fig. \ref{fig:figura_frec1}). The remaining 20 elements (see Fig. \ref{fig:figura_frec2}) are the non-alphabetic chars commonly used in injections. The variables (except the last one) represent the number of matches of the $Payload$ within the query. On the contrary, the last variable is the final ranking with the $level$ value. So, for example a $\vec{x} = \{1, 0, 3, \dots 4\}$ will represent a query with one $union select$, zero $all$, three $and$ and it is classified as high risk (level 4).

\subsection{Classifier and Performance}
\begin{figure*}[!ht]
	\includegraphics[width=0.8\textwidth]{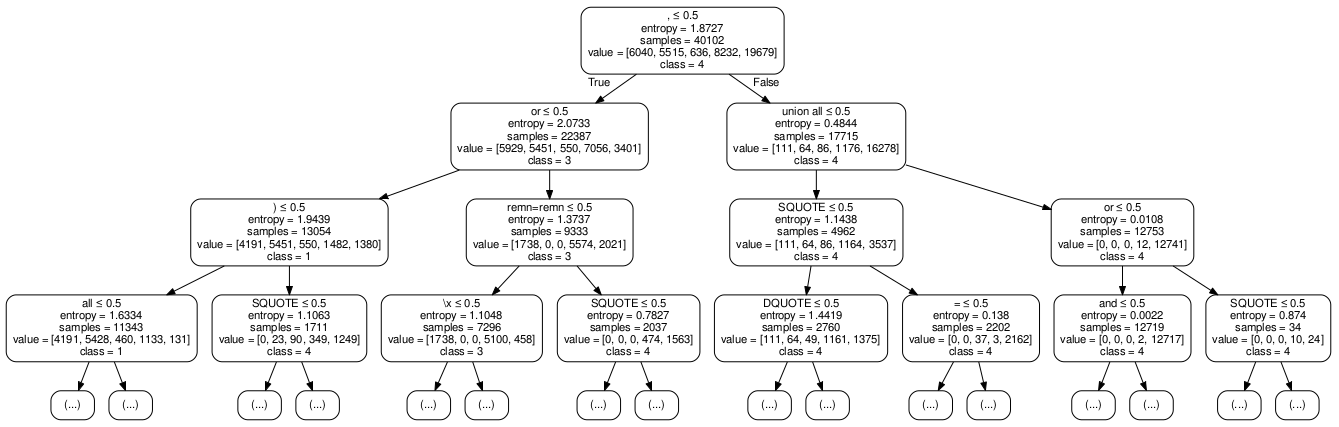}
    \caption{Decision Tree Classifier}
    \label{fig:classifier}
\end{figure*}

The training has been performed using the Python library ``sci-kit-learn'' (a machine learning library) using the corresponding to the file resulting from the classification seen in the previous subsection with a total of 50,000 rows. The classifiers have been trained by two different models known as "Naive Bayes Classifier" and a tree type classifier (see Fig. \ref{fig:classifier}).
The Bayes classifier is popular for the classification of information based on the Bayes theorem for supervised probabilistic classification. Unlike other works, the sample of 50k elements gives a very high effectiveness (the more tests, the higher the probability of classifying correctly). Using a Bayes classifier, an "accuracy score" was obtained after a prediction of $ 0.98 + \alpha / \alpha = \pm 0.1 $. In addition to this, a tree classifier has been used obtaining classification results that are very similar to those of Bayes which indicates that the classification before training is entirely correct. For the "decision tree" type classification, the entropy and a maximum classification level of 10 (tree depth) have been used as criterion element, with a set of tests of 0.3 (leaving 0.7 of training).

\section{Collaborative Work}
\label{sec:col}
\begin{figure*}[hbt!]
    \includegraphics[height=6.5cm,keepaspectratio]{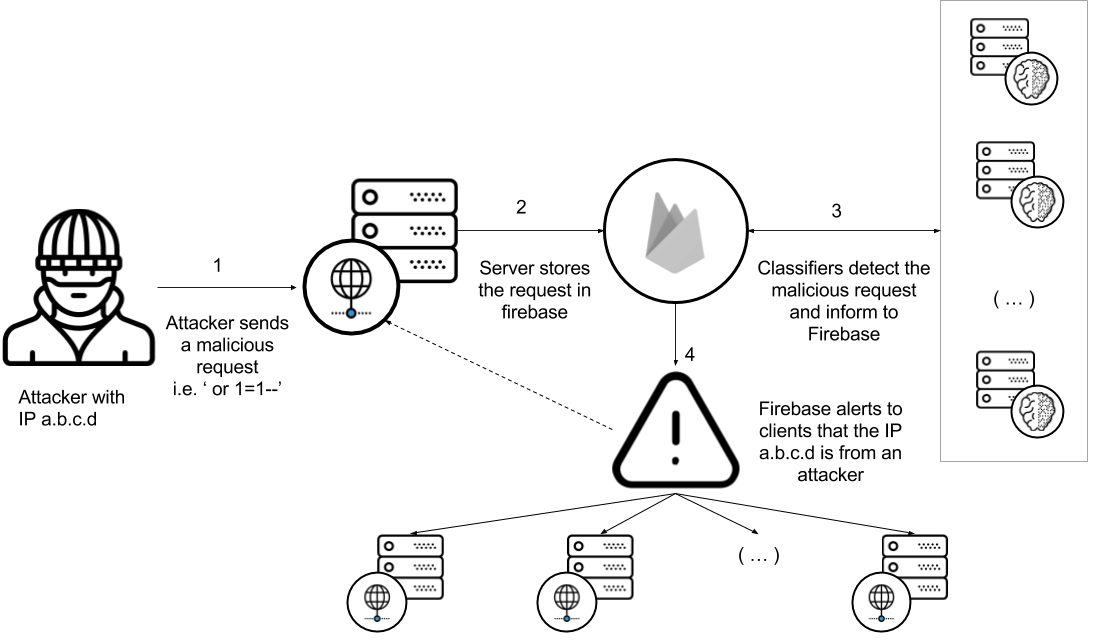}
    \caption{Collaborative machine learning scheme}
    \label{fig:block}
\end{figure*}
In this way, a computer can overcome geographical barriers in order to work in a network of computers (or servers) with the same tasks. For years, the user experience in database engines has been improved, this is the case of Firebase in real time. Firebase is a safe tools suite that allows multiple functionalities such as having a non-SQL database that works by connecting sockets between the client(s) and the server(s). The best of it is that this database has real-time events like listeners that notify of changes when they happens, at this moment, the customers that are listening receive the changes made.

\subsection{Distributed Schema}
When thinking about distributed systems it is common to think about computers that work in remote geographic situations. This concept fits the main idea of the paper which is the creation of a distributed system (see Fig. \ref{fig:block}) for the classification of multiple clients based on the classifier system discussed throughout the paper. The main idea is to create a network where each node can be a client or a server. The reason and the need for a distributed system for the protection of services exposed to the internet is mainly the protection of the data of the clients against hackers. It is understood that a client is one who has a service exposed to the internet (such as a web page, a store application, etc.). The incoming requests are recorded not only in a log but also are stored in the database (in this case, Firebase). The advantage of this is that all the clients that are connected in the network send data to a global-shared log in a machine. As discussed at the beginning of the section, Firebase has a direct communication socket system so when a client saves a record in the query database, Firebase automatically notifies the servers that there are new stored items (new queries). These queries can be analysed by the network servers with different types of classifiers in order to check whether the request is a type of attack or not. In the case of a type of attack, an event would be notified through Firebase to all the connected clients in order that the clients can take the decision to block an intrusion prior to an attack. The steps that the system would perform are described in the following figure \ref{fig:block}. Also, a brief description of each step is included.

\begin{enumerate}
  \item An attacker launches a malicious query to the client that is accessible from internet (in this case, a web page).
  \item The web page stores a record in the shared database that is in the cloud (without knowing yet that it is an attack).
  \item The Firebase engine notifies to any client or server that is listening that there are changes in the database. Besides, servers analyses the request(s) using some different classifiers and notify Firebase of the classification of that query.
  \item Firebase notifies to all listener clients if this query is legitimate or is dangerous.
\end{enumerate}

The advantage of this system is that training times do not affect the clients, but only the servers. Another advantage is also that having a shared log system can have a very large database in order to be able to offer a continuous training in the servers improving the classifier with each stored query. 

In a confusion matrix, the value related to Type I (or false positive) occurs when a request is legitimate and classified as dangerous or erroneously. These cases can be related to queries that include tokens belonging to the SQL language such as ``union all'' (for example: ``union all values contained in multiple lists''). According to our classifier, that query could be malicious due to the containment of those tokens. However, not only does it depends on these instructions for classification but on other parameters for determining whether a query is malicious or not. At present, the probability of detection of a false positive (according to the dataset and training data) is currently close to $\pm 1\%$.

\section{Conclusions}
\label{sec:conc}
This paper presents an improvement of current intrusion detection systems based on the use of a frequency analysis and the previous behavior of one of the most used database audit software, SQL Map. The main result is a positive detection (close to 99 \% positive detection) based on thousands of test and training data (almost 50,000 queries) used to improve the performance of current systems. In addition to this, we have performed a system for the management of intrusion detection by using the above-mentioned tool. As future work, we are working on improving both tools to offer an even higher success rate (including decreasing the number of false positives) and extending the analyses to other database engines used today (based on relational and non-relational models).

\section*{Acknowledgements}
Research supported by the Spanish Ministry of Economy and Competitiveness, the European FEDER Fund, and the CajaCanarias Foundation, under Projects TEC2014-54110-R, RTC-2014-1648-8, MTM2015-69138-REDT and DIG02-INSITU.

\bibliographystyle{IEEEtran}
\bibliography{acmart} 

\end{document}